\documentclass[usenatbib,useAMS]{mn2e}
\usepackage{epsfig}
\usepackage{times}
\usepackage{amsmath}
\usepackage{amssymb}
\usepackage{verbatim}
\usepackage{color}
\usepackage{paralist, tabularx}

\def\gtrsim{\mathrel{\hbox{\rlap{\hbox{\lower3pt\hbox{$\sim$}}}\hbox{\raise2pt\hbox{$>$}}}}}
\def\lesssim{\mathrel{\hbox{\rlap{\hbox{\lower4pt\hbox{$\sim$}}}\hbox{$<$}}}} 
\newcommand{\simgt}%
        {\,\hbox{\lower0.6ex\hbox{$\sim$}\llap{\raise0.6ex\hbox{$>$}}}\,}
\newcommand{\simlt}%
        {\,\hbox{\lower0.6ex\hbox{$\sim$}\llap{\raise0.6ex\hbox{$<$}}}\,}
\newcommand{\msun}{\ensuremath{\mathrm{M}_\odot}}

\title[Helium accretion and SNe~Ia]{The effect of helium accretion
  efficiency on rates of Type~Ia supernovae: double-detonations in
  accreting binaries}
\author[A.~J.~Ruiter et al.]{A. J. Ruiter$^{1}$\thanks{E-mail:
    \texttt{ajr@mpa-garching.mpg.de}},  K. Belczynski$^{2,3}$, S. A. Sim$^{4}$,
  I. R. Seitenzahl$^{5,1}$, D. Kwiatkowski$^{6}$\\
$^{1}$Max-Planck-Institut f\"{u}r Astrophysik, Karl-Schwarzschild-
Str. 1, D-85741 Garching, Germany\\
$^{2}$Astronomical Observatory, University of Warsaw, Al.
            Ujazdowskie 4, 00-478 Warsaw, Poland\\
$^{3}$Center for Gravitational Wave Astronomy, University of Texas at
            Brownsville, Brownsville, TX 78520, USA\\
$^{4}$Astrophysics Research Centre, School of Mathematics and
  Physics, Queen's University Belfast, Belfast BT7 1NN, UK\\
$^{5}$Institut f\"{u}r Theoretische Physik und Astrophysik, Universit\"{a}t W\"{u}rzburg, Am Hubland, D-97074 W\"{u}rzburg, Germany\\
$^{6}$Department of Physics, University of Warsaw, Ho\.{z}a 69, Warsaw, Poland
\\ 
}

\date{Accepted 17 February 2014}

\pagerange{\pageref{firstpage}--\pageref{lastpage}}
\pubyear{}
\begin{document}
\maketitle
\label{firstpage}

\begin{abstract}

The double-detonation explosion scenario 
of Type~Ia supernovae has gained increased support from the SN~Ia 
community as a viable progenitor model, making it a promising 
candidate alongside the well-known single degenerate and double 
degenerate scenarios.  
We present delay times of double-detonation SNe, in
which a sub-Chandrasekhar mass carbon-oxygen white dwarf accretes 
non-dynamically from a helium-rich companion.
One of the main uncertainties in quantifying SN 
rates from double-detonations is the (assumed) retention efficiency of
He-rich matter.  
Therefore, we implement a new prescription 
for the treatment of accretion/accumulation of He-rich 
matter on white dwarfs.  
In addition, we test how the results change depending on 
which criteria are assumed to lead to a detonation in the helium shell. 
In comparing the results 
to our standard case (Ruiter et al.\ 2011), 
we find that regardless of the adopted He accretion prescription,
  the SN rates are reduced by only ${\sim}25$ per cent if 
low-mass He shells ($\lesssim 0.05\, \msun$) are
sufficient to trigger the detonations.
If more massive ($0.1\, \msun$) shells are needed, the rates
decrease by $85$ per cent and the delay time distribution is
significantly changed in the new accretion model -- 
only SNe with prompt ($< 500\, \mathrm{Myr}$)
delay times are produced.  
Since theoretical arguments favour 
low-mass He shells for normal double-detonation SNe, 
we conclude that the rates from double-detonations are likely to be high, 
and should not critically depend on the
adopted prescription for accretion of He. 

\end{abstract}


\section{Introduction}
\label{sec:intro}

It is a widely-accepted view that Type Ia supernovae (SNe~Ia) arise
from the thermonuclear explosion of a white dwarf (WD) star  
(see \citealt{hillebrandt2013a}).  
Until a few years ago,
the favoured progenitor scenario that was said to lead to SNe~Ia was
the single degenerate scenario (SD), by which a carbon-oxygen (CO) WD
accretes from a (probably hydrogen-rich) non-degenerate
companion star, until the WD's central density becomes sufficiently high to
ignite carbon.  Such high densities are likely achieved for CO WDs
that approach the Chandrasekhar mass limit (${\sim} 1.4$ \msun).  The
other well-known progenitor scenario 
is the double degenerate (DD) scenario, in which two
WDs merge. 
Previously, it was expected
that the primary WD had to achieve near-Chandrasekhar mass before
explosion, though it is becoming more clear that 
this is not necessarily the case:  
Recent work has shown that sub-Chandrasekhar mass WD
explosions are successful in synthesizing $^{56}$Ni in sufficient
amounts during violent mergers \citep[see e.g.][]{pakmor2012a}.  

A third progenitor scenario that has recently gained more positive
attention is the double-detonation
scenario, in which a detonation is triggered off-centre in a
sub-Chandrasekhar mass WD following an initial detonation in a He
layer (or `shell') that has been accumulated on the WD surface
\citep[e.g.][]{livne1990a,iben1991a,woosley1994b,livne1995a,fink2010a,townsley2012a,moore2013a}.    
Early studies indicated that this `classic' double-detonation scenario -- where a
CO WD accumulates mass from a He-rich companion that is stably
filling its Roche lobe\footnote{Note that double-detonation explosion
  mechanisms may also be encountered during mergers that proceed
    on dynamical timescales \citep{guillochon2010a,pakmor2013a,shen2013a}.
  Throughout this paper, {\em we do not discuss mergers}, and we refer to `double-detonation' to mean
  systems in which the companion star is filling its Roche lobe and
  transferring matter to the primary WD on a non-dynamical
  timescale.} -- was not a promising SN~Ia scenario since
theoretical spectra and lightcurves did not match those of normal SNe~Ia
\citep{hoeflich1996a,nugent1997a,garcia1999a}.  However, more recent work has shown
that lightcurves, spectra and nucleosynthesis from these explosions
may compare relatively well with observational data
\citep{kromer2010a,woosley2011b}.  The main difference between the recent
studies and those performed in the 1990s is the realization that a
thick (${\sim}0.2$ \msun) He shell is likely not needed for a
detonation.  
With a lower He shell mass, 
there is no longer a significant over-production of Fe-peak elements at 
high velocities, which brings model spectra into better agreement with
SN~Ia observations.   

As discussed in \citet{ruiter2011a}, the double-detonation
model for SNe~Ia is attractive for several
reasons: 
\begin{compactitem}
\item The lack of hydrogen in SN~Ia spectra is a natural result.
\item The range in exploding (primary WD) mass provides a simple, 
  physical parameter that accounts for the observed variety among SN~Ia
  peak-brightness and lightcurve width.
\item Model spectra and lightcurves show potential for 
looking as good as DD and SD model spectra/lightcurves 
when compared with observational data.
\item Predicted rates are high enough to possibly explain a large
  fraction of SNe~Ia, and the delay time distribution (DTD) compares
  well with observational data \citep{ruiter2011a}. 
\end{compactitem}

These criteria are also fulfilled by the violent white dwarf merger
scenario \citep{pakmor2012a}.  Nonetheless, given the
diversity of SNe~Ia, it is likely that more than
one progenitor channel contributes to the observed population.  
Thus, it is clear that further exploration of the
double-detonation scenario is important.  
In this paper, we re-examine the fourth of these bullet
points.  One factor that is expected to 
strongly affect rates of double-detonations is  
the retention efficiency of He material on WD accretors, and so we
test the assumptions involved in the physical treatment of this process 
with new input physics. 
\citet{piersanti2013a} (hereafter P13) 
concluded that the \citet{ruiter2011a} rates of
double-detonation SNe~Ia -- whose progenitors were WDs with 
total masses $>0.9$ \msun\ -- 
are likely overestimated.
When taking into account the thermal response of the
He-accreting WD in long-term evolutionary calculations,
P13 found it unlikely that the CO WD would 
grow substantially in mass during high mass transfer rates, in
contrast to \citet{kato2004a} (hereafter KH04).   
To test this, we have implemented a new prescription for 
the retention efficiency of He-rich matter 
into our binary evolution calculations that 
is based on the study of P13.     
Another factor to consider, as it 
likely affects the explosion masses, is the assumed criteria   
leading to a detonation in the He shell. We test different
conditions for this as well. 

\section{Modelling: old vs. new input physics}
\label{sec:model}

Despite being an integral piece of physics to the understanding of
SNe~Ia and interacting binaries in general, our theoretical
picture of retention efficiency in mass-transferring binaries remains incomplete.\footnote{Although it is important to also consider
  the effect of retention efficiency of hydrogen-rich
  material on SN~Ia progenitors in general \citep{idan2013a}, we do not explore that
  here.  For double-detonation SN~Ia candidates, stable mass
  transfer phases involving hydrogen-rich donors are less important.} 
In order to quantify the total number (and relative
frequency) of SNe~Ia that may arise from the proposed 
formation channels, we must turn to binary population synthesis (BPS)
methods 
\citep[see][for a comprehensive BPS comparison study]{toonen2013a}. 
Various prescriptions for the treatment of mass
accretion have been adopted in different 
BPS codes, and the different parametrizations/prescriptions 
are one of the factors contributing to the 
variability in SN~Ia rate predictions among different groups 
\citep[see][]{mennekens2010a,nelemans2013b,bours2013a}.

\citet{ruiter2011a} adopted the He accretion  
prescription of \citet{kato1999a,kato2004a} and assumed that a 
He shell mass of 0.1
\msun\ was needed to trigger a double-detonation 
\citep[see also][]{belczynski2005a}.  
In addition, 
\citet{ruiter2011a} assumed a double-detonation SN~Ia explosion only occurs if the total WD
mass (CO `core' $+ 0.1$ \msun\ `shell') $\geq 0.9$ \msun.   
In that work, rates and delay times of SN~Ia from several evolutionary channels were calculated with the BPS code {\tt
  StarTrack} \citep{belczynski2002a,belczynski2008a} with three
different parametrizations for the common envelope (CE) phase.  
For our standard model, 
the values $\alpha_{\rm CE}=1$ and
$\lambda=1$ were adopted \citep[see][sect.~3]{ruiter2011a}. Since
our standard model yielded the highest rate of SNe~Ia, in particular
for double-detonation SNe~Ia (Ruiter et al. (2011), table 1), 
we use those results as a benchmark for comparison to
the current study. 

In an accreting binary system, some fraction of material lost 
from the donor remains bound to the accretor.  The value of this
fraction, $\eta$, and exactly how it evolves during binary evolution 
is uncertain.
Nevertheless, if one adopts a recipe prescribing how the amount of
retained matter depends on e.g. the donor mass
transfer rate and the mass of the accreting WD, this can be
incorporated into BPS studies and used to understand how assumptions
about $\eta$ influence predicted properties of a binary population. 
Since larger values
of $\eta$ will generally result in larger CO WD masses, testing different 
treatments for the retention of He-rich matter derived from
different research groups 
is critical in
determining uncertainties in the rates, delay times
and physical properties of SN~Ia progenitors.  This is true in
particular for double-detonation SNe~Ia, but it also has
an effect on the DD and HeR  
(He-rich Chandrasekhar mass WD) scenarios. 
The implications for these other progenitors will 
be discussed in a forthcoming paper (Ruiter et al. in prep.); 
for the current study we focus on double-detonations. 

The response of the accreting WD upon receiving mass depends on
the WD mass and the rate at which mass is being transferred from the
donor \citep[e.g.][]{moll2013a}. 
For very high mass transfer rates (${\sim}10^{-5}$ \msun yr$^{-1}$, 
when mass transfer first begins from a He to a CO
WD)\footnote{Our donor stars consist of He-rich WDs and low-mass
  He-burning stars.  Initial mass transfer rates from these 
`main sequence He stars' are typically low: ${\sim}$few $\times 10^{-8}$ \msun yr$^{-1}$.}, 
the retention efficiency can vary in a wide range.     
When the rate of transfer is higher than the rate of
burning, the transferred He can form a `red giant-like envelope' 
on the WD surface \citep[see][]{nomoto1982a}, and a substantial amount
of material may be lost.  As the mass transfer 
proceeds at lower rate, but is still fairly high (${\sim}$~few $\times
10^{-6}$ \msun yr$^{-1}$),
more of the transferred material is burned and adds to the WD's mass,
and eventually a regime of stable burning can be achieved \citep[][IT89]{iben1989a}. 
As the orbit increases and 
the mass transfer rate drops further, 
burning becomes unstable as the binary enters a flash cycle 
where only some of the transferred matter is accreted, 
the rest being lost from the binary (KH04). 
Lastly, when/if the mass transfer rate drops to a
sufficiently low value (typically  $\dot{M} < 10^{-7}$ \msun yr$^{-1}$),
material accumulates on the WD surface efficiently, but
temperatures are not high enough for He burning. If this He
shell reaches a critical mass, 
the physical conditions in the (degenerate)
He layer may be sufficient to trigger a He flash that 
evolves as a detonation \citep[e.g.][]{taam1980a}.  This first detonation is
then likely to trigger a second detonation closer to the WD centre \citep{fink2010a,shen2013b,moll2013a}. 

{\em Previously-adopted (old) He accretion prescription}:
For accretion of He-rich matter on WDs, the adopted prescription (e.g.
used in Ruiter et al. 2011) is based on 
detailed He flash calculations from KH04.  
They found that for He accretion rates $\gtrsim 10^{-6}$ \msun\ 
yr$^{-1}$, $\eta$ approaches or is equal to 1 (see their fig.~2),
  whereas $\eta$ will have a range of values for lower accretion
  rates. We group the accretion stages described in 
Sect.~\ref{sec:model} into four `regimes' to summarise how the
input physics is treated in our binary evolution calculations (see
KH04 for formulae):\\
i) {\em accretion at high $\dot{M}$}: stable He burning is assumed ($\eta=1$)\\
ii) {\em steady accretion regime}: stable He burning ($\eta=1$)\\ 
iii) {\em helium flash regime}: unstable He burning ($0 < \eta < 1$;
adopted eq. 1-6 from KH04)\\ 
iv) {\em steady accumulation/double-detonation regime}: 
accumulation
of He `shell' ($\eta=1$, no burning). \\ 
The build-up of the He shell that is needed for a
double-detonation to occur is only possible if the binary is evolving
in regime iv. We note that for the KH04 model, regimes i
  and ii are identical in terms of efficiency. While we restrict all
  WD-accretion to be Eddington-limited, assuming 
  $\eta=1$ for high mass transfer rates is likely to over-estimate 
  the amount of mass gained, as
  mentioned in P13.

{\em Newly-adopted He accretion prescription}: 
We incorporate 
an accretion scheme that is based on P13 ($\dot{M}$
vs. $M_{\rm WD}$, their fig.~1).  
Since P13 do not include detailed information 
about accretion efficiencies or formulae, 
we construct a model that assumes the retention
of He-rich matter follows the trends illustrated in P13  
until a more precise treatment becomes available 
(L.~Piersanti, private communication 2012).   
Such a model, though simple, is an important step towards 
quantifying the
effect that different physical treatments for accretion 
have on SN~Ia rates and exploding WD mass.  
We fit the boundaries 
($\dot{M}_{\rm crit}$) that separate 
retention regimes shown in their fig. 1 
using 
$\dot{M}_{\rm crit}=a\mathrm{e}^{bM}$, 
where $a$ and $b$
are the fitted coefficients and $M$ is the mass of the 
accretor (see Table~\ref{tab:coefs}).  
We adopt the following retention regimes:
\\
i) {\em accretion at high $\dot{M}$, so-called `red giant'
  configuration}: 
We assume $\eta=$min($1.0,\dot{M}_{\rm crit}/ \dot{M}$) \\
ii) {\em steady accretion and mild flash regime}: we assume full
efficiency for burning (steady accretion) or accumulation 
(mild flashes), thus $\eta=1$ (see P13)\\
iii) {\em strong flash regime}: we adopt $\eta=0.\overline{3}$ based
on P13 who state that a range between 
$0.11 < \eta < 0.77$ is feasible\\
iv) {\em steady accumulation/double-detonation regime}: 
accumulation
of He `shell' ($\eta=1$, no burning).  

We assume that a double-detonation thermonuclear explosion will 
ensue if a shell of accumulated (unburned) He reaches a critical value.
In one case we assume a value of $0.1 $\msun\ as was adopted in \citet{ruiter2011a}.  
We also explore the case where a double-detonation is 
presumed to occur with a He shell mass of $0.05$ \msun.   
This is a more reasonable assumption given 
recent studies of He accretion with 1D hydrodynamical simulations
in the context of double-detonations  
\citep[][see also \citealt{moore2013a}]{woosley2011b}. 
However, this critical shell mass likely depends on the WD mass 
\citep[see e.g.][]{bildsten2007a}, with shell mass being inversely
proportional to WD `core mass'.  
Therefore, in addition to our constant
shell mass models, we adopt a model that uses CO WD core
mass dependent shells.  
For this, we consider three different shell criteria, since the exact 
conditions that will lead to a He
shell detonation at low $\dot{M}$ are not currently well-constrained.  
The first two cases are based on eq.\ 11a from IT89, which was originally
constructed to estimate ignition shell masses for WDs accreting at
{\em constant} $\dot{M}$. For the first of these we use the $\dot{M}$ value the binary had once it crossed
into regime iv, and for the second we use the {\em instantaneous} value of $\dot{M}$. 
We label these ignition masses $M_{\rm ITc}$ and
$M_{\rm ITi}$, respectively. We additionally consider 
minimum shell masses for dynamical burning 
$M_{\rm SBd}$ from \citet{shen2009a} (their fig. 5, lower curve).
Achieving such a minimum 
shell mass does not necessarily lead to shell ignition, though 
in theory, these masses represent a lower limit on the
detonation shell mass.
If a binary evolving in regime iv
accumulates a He shell exceeding any of the three
aforementioned shell masses, it is assumed to undergo a
double-detonation.  
By considering three estimates for the critical shell mass 
and assuming that explosion occurs as soon as the smallest one 
is achieved, 
we provide an upper limit on rates of double-detonations within this
mass-dependent shell framework. 
As in \citet{ruiter2011a}, we additionally assume
that a SN~Ia only occurs for systems where the primary WD has a {\em total}
mass $\geq 0.9$ \msun.  Our six  
models are labelled as follows: KH04 prescription with
0.1 and 0.05 \msun\ shell, respectively: K0.1, K0.05;
P13 prescription with 0.1 and 0.05 \msun\ shell,
respectively: P0.1, P0.05; core mass dependent shell masses: K-MDS,
P-MDS, respectively.   


\begin{table}
\caption{Coefficients for the 7 exponential functions 
that we fit using fig.\ 1 of P13 as a guide (see text). 
The table data shown represent the critical limits between two
adjacent regimes. Note that $\eta=1$ is assumed for CO WD accretors with
small initial masses ($<0.61$ \msun, see also KH04).}
  \begin{tabular}{lccc}\hline
    regime & WD mass [\msun] & $a$ [\msun\ yr$^{-1}$] & $b$ [\msun$^{-1}$]\\ \hline
    i-ii      & 0.61 - 0.85   &  1.95964598e-08 & 4.93404225\\
    i-ii      & 0.85 - 1.05   & 3.19735998e-08  & 4.35598835\\
    i-ii      & 1.05 - 1.4    &  4.30115846e-07 & 1.88390002\\
   ii-iii     & 0.61 - 1.025  & 1.93277991e-09  & 5.20188685\\
   ii-iii     & 1.025 - 1.4   &  2.65362072e-08 & 2.66212858\\
  iii-iv      & 0.61 - 0.8    & 9.67049947e-10  & 4.29852144\\
  iii-iv      & 0.8 - 1.0     & 9.28070998e-09  & 1.45637761\\
  iii-iv      & 1.0 - 1.4     & 4.0e-8          & 0\\
\hline
\end{tabular}
\label{tab:coefs}
\end{table}

\begin{figure}
  \centering
  \includegraphics[width=8cm]{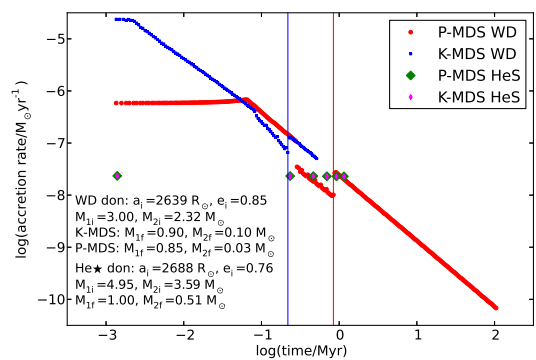}
  \caption{Mass accretion rate as a function of time relative to
    start of mass transfer for double-detonation progenitors 
    in the K-MDS and P-MDS models.  Breaks in the data showing the WD
    donor systems represent transitions between regimes:
e.g.\ from regime ii ($\eta=1$) to regime iii ($\eta<1$) to regime iv
($\eta=1$). 
    The relative times at which regime iv is achieved 
    are indicated by
    vertical lines for the WD donor case (blue; K-MDS and red; P-MDS). 
The system with a helium star donor (HeS) reaches regime iv immediately 
when mass transfer begins for both accretion models. ZAMS parameters
($_{\rm i}$) and masses at explosion ($_{\rm f}$) are shown on the figure. 
}
  \label{fig:accrete}
\end{figure}

In Fig~\ref{fig:accrete} we show examples that lead to a
double-detonation in the K-MDS and the
P-MDS models: a WD donor and a He star donor. 
Both systems undergo two common
envelopes followed by a stable mass transfer phase (plotted). 
The K-MDS WD system initially accretes
with $\eta=1$, while the P-MDS WD system initially
has $\eta=0.015$ (regime i). The K-MDS WD system 
explodes with core and shell masses 0.871
and 0.024 \msun, respectively, when $M_{\rm ITc}$ is achieved. 
The $M_{\rm ITi}$ and $M_{\rm SBd}$ shell masses 
are both within a factor of 2: 0.039 and 0.042
\msun, respectively. 
The P-MDS WD system 
explodes later with core and shell masses 0.781 and
0.066 \msun, respectively, when  
$M_{\rm SBd}$ is achieved. 
The $M_{\rm ITc}$ mass
is very similar: 0.070 \msun, though $M_{\rm ITi}$ is 
unrealistically high: ${\sim}2$ \msun.  This is a
reflection of the fact that eq. 11a from IT89 is a poor estimator of
ignition shell mass for lower WD core masses that require long 
timescales (and therefore large changes in $\dot{M}$) 
to accumulate a sufficient amount of
He. The He-star system undergoes a brief
phase of mass transfer with identical behaviour for both models, 
entering regime iv immediately upon mass transfer. At 
explosion the core and shell masses are 0.966 and 0.032 \msun,
respectively. The shell mass lies in between the dynamical 
mass $M_{\rm SBd}$ (0.028), and the ignition masses $M_{\rm ITc}$ 
(0.033) and $M_{\rm ITi}$ (0.034) \msun, respectively.  


\begin{figure}
  \centering
  \includegraphics[width=8cm]{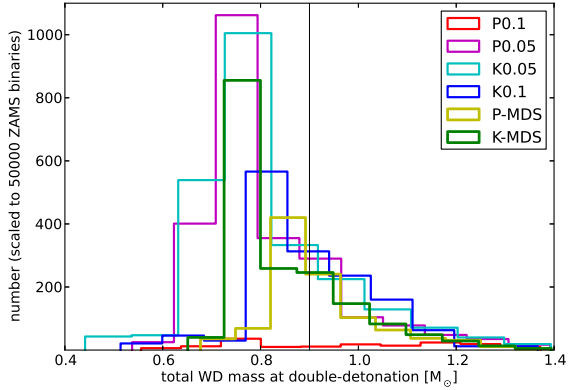}
  \caption{Mass distribution of primary WDs that are
    predicted to undergo double-detonations in {\tt StarTrack} for six models. We show the whole mass range, though only 
    the systems to the right of the vertical black line are 
    likely to explode as SNe~Ia (see text). 
}
  \label{fig:mass}
\end{figure}

\begin{figure}
  \centering
  \includegraphics[width=8cm]{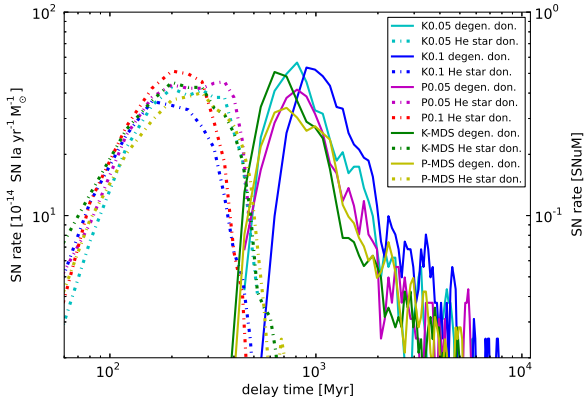}
  \caption{Rates as a function of delay time from
    double-detonations assuming a binary fraction of 70
    per cent.  Only systems that have primary WD masses
    $\geq 0.9$ \msun\
    are shown (see text). The P0.1 model produces only prompt SNe~Ia
    (delay times $<500$ Myr).}
  \label{fig:dtd}
\end{figure}

\section{Results}

In Fig.~\ref{fig:mass} we show  
the mass distribution of WDs that accumulate the
critical shell mass for a double-detonation as predicted by our BPS calculations.
For systems with constant shell mass,  
the 0.05 \msun\ shell models produce a
larger number of events compared to the 0.1 \msun\ shell models that
require twice as much He.  Since we terminate our 
calculation if the donor star mass drops $<$ 0.01 \msun, 
binaries with extremely low-mass donors are excluded from our 
results. For the core mass dependent shell models, lower mass WDs must
accumulate somewhat larger shell masses.  
Consequently, the total WD mass at explosion is systematically 
higher for low mass systems (and slightly lower for high mass systems) 
in the MDS models. 
The peak in K-MDS 
is noticeably higher than the peak in P-MDS due to the assumption of fully efficient 
accretion in regime~i in KH04; the P-MDS 
donor often runs out of mass before any ignition criteria are
reached, and instead the binary evolves as a typical AM CVn system.  
The outcome of double-detonations in low-mass  
CO WDs was explored in \citet{sim2012a}.    
That work has shown that fast transient events can arise from such
systems, with the amount of 
Fe-group and intermediate-mass elements synthesized depending on 
the exact nature of the explosion mechanism.  In any case, the
lightcurves will be fainter and faster-declining than normal SNe~Ia.
Here, we are interested in candidates for SNe Ia of normal brightness.
For this reason, we assume -- as in \citet{ruiter2011a} -- that a
double-detonation SN~Ia only arises in primary WDs of total mass $\geq
0.9$ \msun.
Such an explosion is likely to yield a $^{56}$Ni mass that is around the 
lower limit of observationally-inferred 
$^{56}$Ni masses \citep{sim2010a,ruiter2013a}. 
Though the core and shell masses will have an effect on the
  resulting spectral signature, to first order the total mass of $^{56}$Ni synthesized in a
double-detonation is fixed by the \emph{total} mass of the primary WD.

\begin{table}
\begin{center} 
  \caption{Second column shows relative occurrence rates over a Hubble time, last three
    columns show donor types (by per cent) 
of double-detonation SNe~Ia.  We only list the statistics for events
where the total exploding WD mass is $\ge 0.9$
    \msun.  
}
  \begin{tabular}{l|c|c|c|c}\hline
    model & Rel.~frac. & He WD & Hyb WD & He-star \\ \hline
    K0.1  & 1 & 79 & 10 & 11\\
    K0.05 & 0.92 & 74 & 9 & 17\\
    P0.1  & 0.15 & 0 &  1 & 99\\
    P0.05 & 0.79 & 67 & 10 & 23\\
    K-MDS & 0.76 & 64 & 12 & 24\\
    P-MDS & 0.73 & 68 &  8 & 24\\
    
  \hline
  \end{tabular}
\label{tab:rates}
  \end{center}
\end{table}

The main difference between the two different 
accretion schemes (KH04 and P13) is that KH04 is more favourable for building up the mass of the 
WD, specifically within regime i.  
In addition, the $\eta$ values
achieved during regime iii are generally higher 
in the KH04 models. Consequently, these systems enter the double-detonation
regime with more massive binary components. 

We find that the DTD of double-detonation SNe~Ia is
significantly altered from that of \citet{ruiter2011a} when the 
P13 retention efficiency is adopted and the WD is required to
accumulate a $0.1$ \msun\ He shell (see
Fig.~\ref{fig:dtd}).  The reason has to do with the nature of the
progenitors: they all involve relatively massive donors -- no
He WDs (see Table~\ref{tab:rates}).  
The only double-detonation SN~Ia systems found in the
P0.1 model are those with either He-burning star donors or
(rarely) `hybrid'
WD donors that consist of a CO-core and a He-rich mantle. 
During mass transfer, more matter is lost from the
binary in the P0.1 model 
and the He WD donors run out of matter before the 
critical shell mass is reached. 
Thus, He WD $+$ CO WD binaries cannot make double-detonation
progenitors in P0.1.  
This is the reason for the significant
decrease (by a factor of ${\sim}7$) in the rates of double-detonation
systems in this model compared to \citet{ruiter2011a} (see
Table~\ref{tab:rates}, where the K0.1 model is the one comparable to
the standard results of \citealt{ruiter2011a}).  
However, this decrease
is mitigated if we allow for double-detonations in which a
smaller amount of accumulated He is required, as is the case for
the P0.05, K0.05, P-MDS and K-MDS models.  For the
MDS models, if each ignition shell criterion is considered separately
(rather than choosing the lowest mass), the rates for P13 do
not change for $M_{\rm ITc}$ or $M_{\rm SBd}$, though they drop by $60$ per
cent for $M_{\rm ITi}$. For KH04 the rates do not change for 
$M_{\rm ITc}$, they drop by $20$ per cent for $M_{\rm SBd}$,
and they drop by $40$ per cent for $M_{\rm ITi}$.

\section{Summary}

We have compared rates of double-detonation SNe~Ia arising 
from sub-Chandrasekhar mass CO WDs accreting He-rich matter on
  non-dynamical timescales for
two prescriptions for He retention efficiency.
In addition, we have
tested the prescriptions assuming different critical values for
accumulated He shell mass above which a double-detonation is
presumed to occur: constant shell masses as well as
  CO WD core mass dependent shell masses. 

If a thick ($0.1$ \msun) shell of He is a necessary
condition to achieve a double-detonation SN~Ia, then 
most events will have He-star donors and should be found among young stellar
populations if our newly-adopted retention efficiency prescription (P13) is
assumed.  
This finding is in stark contrast to the results
of \citet{ruiter2011a}, who found that most 
double-detonations will arise from CO WDs accreting from He
WD donors.  
If only thin He shells 
are required, then it will be difficult to disentangle 
progenitor evolution based on delay
time alone, regardless of the assumed mass-retention model.  However, 
the assumed mass-retention model should not significantly affect the
expected rates.  

In contrast to older models that assumed thick shells, 
recent models indicate that thin He shells 
produce observables that agree fairly well with observations 
\citep[e.g.][]{kromer2010a,woosley2011b}. 
This is particularly true for 
double-detonations leading
to `normal' SNe~Ia that call for fairly massive CO
WDs \citep[${\gtrsim} 1$ \msun, see also][]{piro2013a} and 
thus likely require small He shells. 
Understanding the mass dependence of the 
detonating shell is a complex problem. 
Here, we have explored a range of possibilities to estimate the 
WD explosion mass (and rate) by including detonation and ignition shell 
calculations based on core mass and accretion rate. 
Such models (K-MDS and P-MDS) are more realistic than 
assuming a constant shell mass. However, it turns out that the assumed 
ignition criterion is, to first order, not of 
crucial importance if the critical shell mass is low (${\lesssim} 0.05$
\msun): in this case the total rate of double-detonations 
remains high.


\section*{Acknowledgments}
 
The authors thank the anonymous referee for suggestions that improved the manuscript.  
AJR thanks L. Piersanti, L. Yungelson, K. Shen, H. Ritter,
W. Hillebrandt and S. Woosley for discussion. 
KB acknowledges partial support from the Polish Science Foundation
under the Master 2013 program and Polish NCN grant SONATA BIS 2.
IRS was funded by the DFG through
the graduate school GRK~1147.
WebPlotDigitizer was used for some data extraction.  


\bibliography{ashbibz}
\bibliographystyle{mn2e}

\label{lastpage}

\end{document}